\documentclass[12pt]{article}

\begin{document}
\input psfig.sty

\bibliographystyle{unsrt} 


\def\Journal#1#2#3#4{{#1} {\bf #2}, #3 (#4)}

\def\NCA{\it Nuovo Cimento}
\def\NIM{\it Nucl. Instrum. Methods}
\def\NIMA{{\it Nucl. Instrum. Methods} A}
\def\NPB{{\it Nucl. Phys.} B}
\def\PLB{{\it Phys. Lett.}  B}
\def\PRL{\it Phys. Rev. Lett.}
\def\PRD{{\it Phys. Rev.} D}
\def\ZPC{{\it Z. Phys.} C}

\def\st{\scriptstyle}
\def\sst{\scriptscriptstyle}
\def\mco{\multicolumn}
\def\epp{\epsilon^{\prime}}
\def\vep{\varepsilon}
\def\ra{\rightarrow}
\def\ppg{\pi^+\pi^-\gamma}
\def\vp{{\bf p}}
\def\ko{K^0}
\def\kb{\bar{K^0}}
\def\al{\alpha}
\def\ab{\bar{\alpha}}
\def\be{\begin{equation}}
\def\ee{\end{equation}}
\def\bea{\begin{eqnarray}}
\def\eea{\end{eqnarray}}
\def\CPbar{\hbox{{\rm CP}\hskip-1.80em{/}}}

\def\beqn{\begin{equation}}
\def\eeqn{\end{equation}}
\def\beqna{\begin{eqnarray}}
\def\eeqna{\end{eqnarray}}


\title{The Nucleon's Mirror Image: \\
Revealing the Strange and Unexpected}
\author{R. D. McKeown$^{a\dag}$ and M.J.
Ramsey-Musolf$^{a\dag\dag}$\footnote{On leave from Department of Physics,
University of Connecticut, Storrs, CT  06269  USA}
\\
\\
$^a$W. K. Kellogg Radiation Laboratory,\\
California Institute of Technology, \\
Pasadena, CA 91125, USA, \\
\\
\\
$^\dag$E-Mail: bmck@krl.caltech.edu\\
$^{\dag\dag}$E-Mail: mjrm@krl.caltech.edu}

\maketitle

\begin{abstract}

A remarkably successful program of parity-violating
electron scattering experiments is providing new insight into the structure of the nucleon.
Measurement of the vector form factors enables a definitive study of potential strange quark-antiquark contributions
to the electromagnetic structure such as the magnetic moment and charge distribution.
Recent experimental results have already indicated that effects of strangeness are much smaller than 
theoretically expected. In addition, the neutral axial form factor appears to display substantial
corrections as one might expect from an anapole effect.

\end{abstract}

\newpage

Over the last two decades, nuclear and particle physicists have made great strides in
understanding the structure of hadronic matter in terms of the underlying degrees of
freedom associated with Quantum Chromodynamics (QCD). Deep inelastic scattering utilizes 
the electroweak
interaction at large momentum transfers to determine the structure functions of nucleons
associated with
the fundamental constituents, quarks and gluons. These structure functions, and their
consequent evolution in momentum-transfer, 
are theoretically well-defined and interpretable in the context of quantum field theory.
They have been studied with increasing precision over the last 30 years, including remarkable
recent progress in measurements of spin-dependence.

An intriguing aspect of the quark structure of the nucleon is the apparent presence of strange
quark-antiquark ($\bar s s$) pairs. Traditional quark models 
rather successfully describe the nucleon in terms only of
up- and down-flavored quarks. However, since there is no
selection rule forbidding the creation of $\bar s s$ pairs by gluons such
quantum fluctuations should certainly be present at some level. Thus it is perhaps not
surprising that deep inelastic neutrino scattering experiments indicate that 
the $s$ and $\bar s$
each carry about 2\% of the nucleon momentum \cite{bazarko}.
The recent measurements of spin-dependent structure functions have motivated additional
interest in the role of $\bar s s$ pairs. The observed strong violation of the Ellis-Jaffe
sum rule \cite{ji01}
generated a re-examination of the assumption \cite{ellis}
that $\bar s s$ pairs do not significantly
contribute to the quark spin structure of the nucleon. 
These experiments are now interpreted, in a more complete analysis, as 
evidence that angular momentum contributions other than quark helicities are
responsible for nucleon spin (including gluon angular momentum and orbital angular
momentum of the quarks). Nevertheless these studies indicate that, 
of the 30\% of spin carried by quark helicities, the $\bar s s $ pairs have a 
large influence (perhaps 1/3 to 1/2). Unfortunately, $SU(3)$ flavor violating effects 
introduce uncertainty at this level which significantly
diminishes the reliability of this conclusion.

This program has left us in the situation where we know the strange quarks are present, we detect
their presence in deep inelastic scattering processes, but their role in traditional static properties of the nucleon (such as
mass, spin, charge, and magnetism) is still not understood. 
Additional studies of the $\pi$-nucleon sigma
term have indicated that $\bar s s $ pairs contribute up to 1/3 of the mass of the nucleon, but
there are large uncertainties in such an analysis \cite{gasser}. 
Thus, the suggestion of Kaplan and Manohar
\cite{kaplan} that neutral weak form factor measurements could facilitate determination of
$\bar s s $ contributions to the nucleon electromagnetic form factors generated much interest
and led to the proposal that parity-violating electron scattering would be an effective
method to perform such measurements \cite{bmck89,beck89}.

The program of parity-violating electron scattering has now produced its first results, putting it
well on the way towards a quantitative statement about the relative importance of up, down, and strange quarks in the
nucleon's electromagnetic properties. It should be emphasized that -- in contrast to the situation regarding 
flavor content of the nucleon mass and spin -- the determination of $s\bar s$ vector current
properties is free from the kinds of theoretical ambiguities which complicate the flavor decomposition of
these other nucleon properties. Indeed, the set
of highly-precise parity-violation measurements at MIT-Bates, Jefferson Lab, and
Mainz\cite{hasty,spayde,HAPPEX2,g0,pva4} -- coupled with the careful theoretical delineation of various
contributions to parity-violating observables\cite{musolf94a,mus92a} -- will provide a definitive and
theoretically clean study of the low-energy quark and gluon structure of the nucleon.

In the process of pursuing an experimental program of parity-violating electron scattering
it has been realized \cite{musolf90} that higher order electroweak corrections must be taken
into account. Although the corrections to the vector form factors of interest are small and
under good theoretical control, the neutral weak axial form factor potentially contains substantial
contributions from processes that are not well understood theoretically. This aspect
of the parity-violating electron-nucleon interaction includes anapole effects and other
electroweak corrections that are relevant to precision studies of beta decay and
atomic parity violation. Indeed, recent results from the SAMPLE experiment \cite{hasty} indicate
that these axial corrections are substantial and even larger than estimated by theorists.

While the subject of the nucleon's strangeness content is broadly familiar, the topic of axial radiative
corrections -- and the corresponding significance of the SAMPLE result -- is less so. In this note, we therefore
give a brief overview of the subject in hopes of generating a better appreciation of this new and important area of
research.

\begin{figure}
\centerline{\psfig{figure=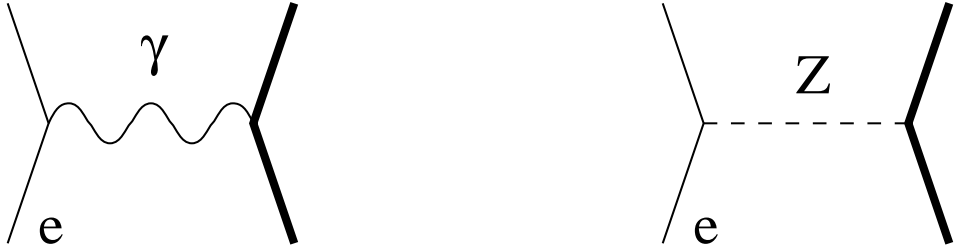,angle=270,width=5.0in}}
\caption {The amplitudes relevant to parity-violating electron
scattering. The dominant parity-violating effects arise from
the interference of these two  
amplitudes.}
\label{fig:amplitudes}
\end{figure}

\section{Parity Violation in Elastic Electron-Nucleon Scattering}

As shown in Figure 1, the lowest-order contribution to the parity-violating
$e$-$N$ interaction is associated with the interference of $Z$-exchange with the dominant
electromagnetic amplitude. The parity-violating helicity-dependent asymmetry for 
elastic electron-proton scattering can be written \cite{musolf94a}:
\begin{eqnarray}
A &=& \left[- G_F Q^2 \over 4 \sqrt{2} \pi \alpha \right]
{{
\varepsilon G^{\gamma}_{{E}} G^{Z}_
{{E}} + \tau G^{\gamma}_{{M}}
G^{Z}_{{M}} - (1-4 \sin^2 \theta_W ) 
\varepsilon^{\prime} G^{\gamma}_{{M}} G^{e}_{A}}   \over
{\varepsilon (G^\gamma_
{{E}})^2 + \tau (G_{{M}})^2}} \\
 &\equiv & -{G_FQ^2\over 4 \sqrt{2} \pi\alpha}\times {{\cal N}\over 
    {\cal D}}\>
\end{eqnarray}
where $\tau$, $\varepsilon $, and $\varepsilon^{\prime} $
are kinematic quantities, $Q^2>0$ is the four-momentum transfer, and
$\theta$ is the laboratory electron scattering angle. This asymmetry represents
the fractional change in cross section for left- vs. right- handed incident electrons, and
is generally quite small
\begin{equation}
A \sim 10^{-4} Q^2
\end{equation}
where $Q^2$ is expressed in units of (GeV/$c$)${}^2$. Thus the experiments are quite challenging.

The quantities $G_E^\gamma$, $G_M^\gamma$, $G_E^Z$, and $G_M^Z$ are 
the vector form factors of the nucleon associated with $\gamma$-
and $Z$-exchange. The neutral weak $N$-$Z$ interaction also involves an axial vector
coupling $G_A^e$ in the third term of the numerator in Eqn.(1).
The lowest-order $Z$-exchange process is responsible for the $1-4 \sin^2 
\theta_W$ factor that appears in this expression and thus
higher order processes can contribute significantly to this term \cite{musolf90,musolf94a}. 
These processes include effects not present in neutrino scattering, such as
anapole effects and other electroweak radiative corrections as discussed below.

It is also useful to consider parity-violating quasielastic
scattering from nuclear targets, particularly deuterium \cite{comment91}.
This provides additional useful information on the 
axial vector form factor contributions. 
For a nucleus with $Z$ protons and $N$ neutrons the asymmetry
can be written in the simple form (ignoring final state interactions
and other nuclear corrections):
\begin{equation}
A_{\rm nuc} = -{G_FQ^2\over 4 \sqrt{2}\pi\alpha}\times 
   {N{\cal N}_n + Z{\cal N}_p \over
    N{\cal D}_n + Z{\cal D}_p}
\end{equation}
where ${\cal N}_p$ (${\cal N}_n$) is the numerator expression and
${\cal D}_p$ (${\cal D}_n$) the denominator (from Eqns. 1 and 2) 
for the proton (neutron), respectively.

\section{Strangeness and the Vector Form Factors}

The neutral weak vector form factors $G_E^Z$ and $G_M^Z$ appearing in Eqn. 1
contain information related to
the desired strange quark-antiquark contributions to the charge and magnetization distributions
of the nucleon. The flavor structure of the electroweak couplings and isospin symmetry of
the nucleon imply the relations
\begin{equation}
G_{E,M}^{s}
 = (1 - 4 \sin^2 \theta_W) G_{E,M}^{\gamma,p}  
		- G_{E,M}^{\gamma,n} - G_{E,M}^{Z,p} \> .
\end{equation}
Thus measurement of the neutral weak form factors $G_{E,M}^{Z,p}$
can unambiguously determine the strange form factors $G_{E,M}^{s}$.

One traditionally defines
\begin{equation}
\mu_s \equiv G_M^s(Q^2=0)
\end{equation}
as the strange magnetic moment of the nucleon. Since the 
nucleon has no net strangeness, we find $G_E^s(Q^2=0) = 0$. However, 
one can express the slope of $G_E^s$ at $Q^2=0$
in the usual fashion in terms
of a ``strangeness radius'' $r_s$
\begin{equation}
r^2_s\equiv -6\left[dG_E^s/dQ^2\right]_{Q^2=0} \> .
\end{equation}

A variety of theoretical methods have been employed in efforts to
compute the form factors $G_{E,M}^s(Q^2)$ (or often just the
quantities $\mu_s$ and $r_s$). 
Typically one may consider the
fluctuation of the nucleon into strange particles ({\em e.g.}, a $K$-meson and hyperon)
or the fluctuation of the virtual boson (photon
or $Z$) into a $\phi$ meson. The physical
separation of the $s$ and $\bar s$ in such processes or the
production of an $s\bar s$ pair in a spin triplet leads to non-zero values of
$G_{E,M}^s(Q^2)$.  The numerical results of many theoretical treatments 
\cite{annrev01} vary considerably,
but generally one obtains a value for $\mu_s \sim \pm 0.5$ (nuclear magnetons) and $r_s^2 \sim
\pm 0.2$ fm${}^2$.

\section{Neutral Weak Axial Form Factor}

As noted above, the parity-violating interaction of electrons
with nucleons also involves an axial vector coupling to the nucleon, $G_A^e$.
This term in the parity-violating asymmetry contains several
effects beyond the leading order $Z$- exchange which can only be differentiated 
in theoretical calculations.  Nevertheless, it is important to bear in mind that the
{\it experimentally observable} quantities are well-defined and unambiguous. 
To this end, we define the neutral weak axial form factors as observed in
neutrino scattering, $G_A^\nu$, and the corresponding quantity $G_A^e$, as 
indicated in the expression Eqn. (1). In the following, we discuss the 
relationship of each of these observables to nucleon beta decay, 
$W$- and $Z$-exchange, and higher order effects such as the anapole moment.

The standard electroweak model relates the axial coupling, $G_A$, measured
in the charged current process (such as neutron beta decay) 
to the neutral current process (such as elastic neutrino scattering). For the case
of neutrino scattering, the interpretation of $G_A^\nu$ is simplified because
the neutrino has no (to lowest order) electromagnetic interaction.
However, due to the effect of $\bar s s $ pairs in generating the
isoscalar neutral weak form factor, in lowest order 
we have the relation
\begin{equation}
G_A^\nu = -G_A \tau_3 + \Delta s \> .
\label{eqn:gaz}
\end{equation}
Here, $\Delta s $ is the same, scale-dependent quantity that appears in the treatment of
spin-dependent deep inelastic scattering\footnote{The scale appropriate for the analysis of
deep inelastic data is considerably higher than for low-energy neutrino reactions. The evolution of
$\Delta s$ between these two scales has not been well-established.}.
$G_A (Q^2 =0) = 1.2601 \pm 0.0025$ is determined in neutron beta decay,
and the $Q^2$ dependence is measured in charged current neutrino scattering
to be
\begin{equation}
G_A (Q^2) = {{G_A (Q^2 =0)} \over {(1+ {Q^2 \over M_A^2})^2 }}
\end{equation}
with $M_A \simeq 1.05$ GeV. In higher order, $G_A^\nu$ also contains
contributions from electroweak radiative corrections leading to the modified expression
\begin{equation}
G_A^\nu = -G_A \tau_3 + \Delta s \>  + R_\nu
\end{equation}
where $R_\nu$ represents the radiative corrections which are of order $\alpha$ as one would
expect \cite{sirlin,musolf90}.

\begin{figure}
\centerline{\psfig{figure=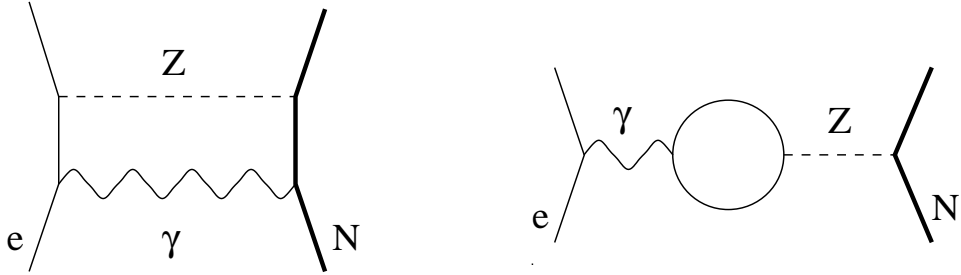,width=5.0in}}
\caption {Examples of amplitudes contributing to the electroweak radiative
corrections $R_e$ (``$\gamma-Z$ box'' on the left) and anapole corrections
(``$\gamma-Z$ mixing'' on the right). Note that these do not contribute to neutrino
scattering corrections $R_\nu$.}
\label{fig:radcorr}
\end{figure}

As is evident in Eqn 1, for parity-violating electron scattering the 
neutral weak axial form factor corresponding
to tree-level $Z$-exchange is multiplied by the small vector coupling of the
electron and $Z$, $|g_V^e|=1 - 4 \sin^2 \theta_W \ll 1$. This suppression of the leading
amplitude increases the relative importance of higher order electroweak contributions, including
those generated by  anapole effects and other electroweak radiative corrections. In particular,
processes involving $\gamma$-exchange between the electron and proton, as illustrated in 
Figure~\ref{fig:radcorr}, do not contain the $g_V^e$ suppression factor, and they can produce sizeable
corrections to the leading order amplitude. To make this situation evident, we may write
\begin{equation}
\label{eq:gaedef}
G_A^e = G_A^Z + \eta F_A + R_e
\end{equation}
where 
\begin{equation}
\eta = {{8 \pi \sqrt{2} \alpha} \over {1 - 4 \sin^2 \theta_W}} = 3.45,
\end{equation}
$G_A^Z =  -G_A \tau_3 + \Delta s$ (as in Eqn.~\ref{eqn:gaz}), 
$F_A$ is the nucleon anapole form factor (defined below), and 
$R_e$ are other electroweak radiative corrections. 

As discussed in \cite{musolf90,musolf91}, the separation of $F_A$ and $R_e$ is 
actually a theoretical issue and dependent upon the choice of gauge. No set of experiments
can yield a separate determination of either quantity; only the sum of terms in
Eq. \ref{eq:gaedef} is measurable. For purposes of intuition, however,
it is useful to consider these quantities separately. Indeed, this situation is similar to
the one encountered in the consideration of the gluon helicity $\Delta G$ that
appears in spin-dependent deep inelastic scattering. That quantity is also
gauge-dependent, but is naturally associated with the gluon helicity in a 
particular gauge, the ``axial gauge''. This gauge dependent quantity is
commonly referred to as the gluon spin structure function \cite{ji01}
and is quoted in relation to a variety of experimental observables.

The anapole moment has been traditionally defined as the effective
parity-violating coupling between a photon and a nucleon \cite{zeldovich}. 
It appears as an additional term in the $\gamma N$ interaction when one includes the possibility that
parity is not strictly conserved \cite{musolf91,musolfphd}:
\begin{eqnarray}
\varepsilon_\mu\langle N| \hat V^\mu_\gamma  | N \rangle &\equiv& -e \varepsilon_\mu\, {\bar u_N} (p^\prime)
\{ F_1 \gamma^\mu - {i \over {2 m}} F_2 \sigma^{\mu \nu} q_\nu \nonumber \\
&+& F_A \, 
[ G_F (q^2 \gamma^\mu - q^\nu \gamma_\nu q^\mu ) \gamma^5 ]
\} u(p)\ \ \ ,
\end{eqnarray}
where $\varepsilon_\mu$ is the photon polarization vector. 
The anapole term, proportional to $F_A$, vanishes for real photons, which have $\varepsilon\cdot q=0$
and $q^2=0$. It contributes only to parity-violating processes involving virtual photons, where its
effect is experimentally indistinguishable from other virtual processes, such as the $Z-\gamma$ box
diagram in Fig. \ref{fig:radcorr}.
Note also that our normalization of $F_A$ differs from that used in the
atomic physics literature by a factor of $m^2 G_F$ with the
result that we expect the value of $F_A$ could be of order unity.
Thus, $F_A$ could indeed provide a substantial contribution to $G_A^e$ (see
Eqn. 11).

The anapole moment has also been considered previously in atomic parity violation
experiments. Its definition is analogous to that in Eqn. 13 above, except that
it is now a form factor of the atomic nucleus (which may involve many nucleons).
In that case, it is expected that the anapole moment will be dominated by
many-body weak interaction effects in the nucleus \cite{haxton}. 
A value for the anapole moment of the Cesium atom has recently been reported \cite{weiman}.

In the case of parity-violating $eN$ scattering, 
the anapole type effects associated with the ``$\gamma - Z$ mixing'' (Fig. \ref{fig:radcorr})
amplitudes are, in fact, the dominant correction\cite{marciano,musolf90,musolf91}. Contributions
to this amplitude from gauge boson, charged lepton, and heavy quark loops can be computed reliably 
in perturbation theory. More care is required, however, in treating non-perturbative strong interaction
effects in light quark loops.
It is conventional to estimate these effects using a dispersive treatment of
$\sigma(e^+e^-\to{\mbox{hadrons}})$ data and flavor SU(3) arguments. This approach may be appropriate
for purely leptonic scattering, but it  does not give a complete treatment for a proton target. For example, the
impact of strong interactions between the virtual quarks in the $Z-\gamma$ mixing loops and those in the target
hadron are not included in the dispersion relation analysis. 
Contributions to $F_A$ involving $Z$- and $W$-exchange between the nucleon's quark constituents have been
estimated using various methods \cite{musolf90,mus00,bira1,riska00} and 
found to be relatively small. In particular, the chiral perturbation theory treatments of Refs.
\cite{musolf90,mus00} include estimates of short distance contributions which may partially account for
strong interactions between the $Z-\gamma$ mixing loops and target quarks. Nevertheless, the
appropriate matching of the dispersion relation and effective field theory treatments remains an open
theoretical question. 

Additional theoretical issues arise when considering the $Z-\gamma$ box contributions to $R_e$. 
The intermediate hadronic state is generally assumed to be a nucleon in all previous calculations
\cite{marciano,musolfphd}. It is possible, however, that there are significant contributions associated with
intermediate
$\Delta$ states and other nucleonic excitations. This consideration applies equally to the 
corrections to neutron and nuclear beta decay (``$\gamma-W$ box'' contributions), neutrinoless
$\beta\beta$-decay (``$W-W$ box" diagrams), and nuclear spin-dependent effects in atomic parity violation
($Z-\gamma$ box diagrams). In principle, a similar statement also applies to precision
studies of neutral weak vector form factors in parity-violating electron scattering and determinations of the
\lq\lq weak charge" in atomic parity-violation. In the case of vector hadronic amplitudes, however, the
$Z-\gamma$ box contributions are suppressed by $g_V^e$\cite{marciano,musolfphd,mus00b}, and the 
corrections are more reliable.
The issue is more serious for $\beta$- and $0\nu\beta\beta$-decay, which provide tests of the
unitarity of the CKM matrix and lower bounds on the mass of heavy, Majorana neutrinos, respectively, as well
as for nuclear spin-dependent effects in atomic parity violation used to probe the nuclear anapole moment. 
Thus, achieving a better understanding of electroweak radiative corrections on $G_A^e$ could have far-reaching
consequences for other precision, electroweak studies. Consequently, the study of the anapole contributions and
other corrections to $G_A^e$ is presently an active area of experimental and
theoretical investigation.

The theoretical issues pertaining to $G_A^e$ have one additional implication. Because in Eq. \ref{eq:gaedef}
$\Delta s$ appears in a linear combination involving the $\eta F_A$ and $R_e$ terms, 
parity-violating electron scattering is not well suited to a determination of $\Delta s$\cite{musolf90}. 
As an alternative, it may be possible to achieve a cleaner determination of $G_A^s$
in low energy neutrino scattering \cite{garvey92}, where the axial vector term
dominates the cross section and the radiative corrections are under better
control. (There is no $g_V^e$-suppression of the leading $Z$-exchange
amplitude, and diagrams such as those in
Figure~\ref{fig:radcorr} involving a photon exchange do not
contribute to  neutrino scattering.)

\begin{figure}
\centerline{\psfig{figure=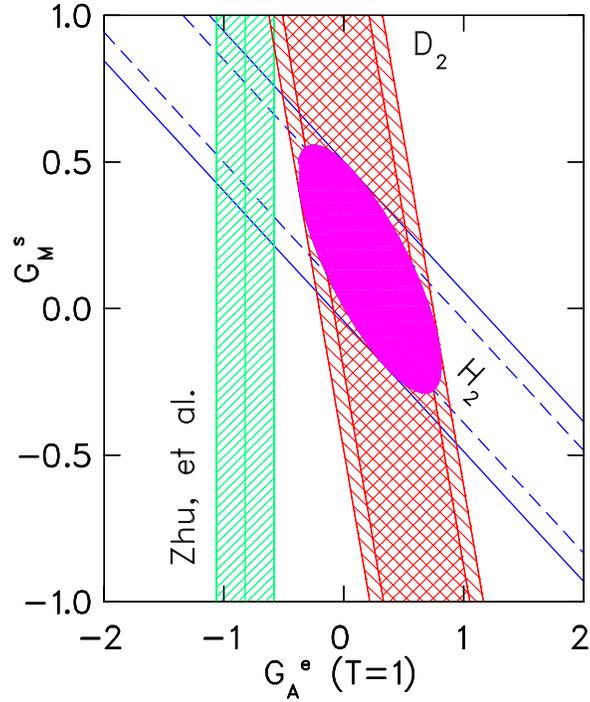,width=3.0in}}
\caption {Combined analysis of the data from the
two SAMPLE measurements. The two error bands from the hydrogen
experiment \cite{spayde} and the  deuterium
experiment \cite{hasty} are indicated. The inner hatched region includes the
statistical error and the outer represents the systematic uncertainty
added in quadrature.  Also plotted is the calculated
isovector axial $e$-$N$ form factor $G_A^e (T=1)$
obtained by using the anapole form factor and radiative 
corrections by Zhu {\it et al.} \cite{mus00}. The typical theoretical prediction
that $G_M^s \sim -0.3$ \cite{annrev01} coupled with the calculation of $G_A^e (T=1)$ is 
substantially ruled out by the experimental data.}
\label{fig:sampleResult}
\end{figure}

\section{Experimental Results}

Three experimental teams have obtained data on parity-violating elastic electron-nucleon
scattering thus far. The first is the SAMPLE experiment at MIT/Bates, which measures the
asymmetry at backward angles from both the proton and deuteron at low $Q^2 = 0.1$ (GeV/$c)^2$.
Those measurements are sensitive to the strange magnetic form factor $G_M^s$ and the
isovector axial form factor $G_A^e (T=1)$, and the results \cite{hasty} are shown in Figure 3.
The measurements indicate
that the magnetic strangeness is small
\begin{equation}
G_M^s (Q^2 = 0.1) = 0.14 \pm 0.29 \pm 0.31 
\end{equation}
and consistent with an absence of strange quarks.
We can correct this value for the calculated $Q^2$ dependence
of $G_M^s$ using SU(3) chiral perturbation theory \cite{hemmert} to obtain a result for the strange magnetic moment:
\begin{equation}
\mu_s = 0.01 \pm 0.29 \pm 0.31 \pm 0.07
\end{equation}
where the third uncertainty accounts for the additional uncertainty
associated with the theoretical extrapolation to $Q^2 = 0 $. An interesting theoretical question is whether
SU(3) chiral perturbation theory provides a reliable guide as to this $Q^2$-dependence (see,
{\em e.g.}, Ref. \cite{hammer}). Future measurements performed at other values of momentum transfer should
provide an answer. 

In addition, the SAMPLE experimental result 
indicates that the substantial modifications of $G_A^e$
predicted in \cite{musolf90} are present, but probably with an even larger
magnitude than quoted in that work. 
It therefore appears that the neutral axial form factor determined
in electron scattering is substantially modified from the tree-level
$Z$-exchange amplitude (as determined in elastic $\nu$-$p$ scattering).
Assuming the calculated small isoscalar axial corrections
are not grossly inaccurate, the isovector axial form factor can be
determined from the SAMPLE results
\begin{equation}
G_A^e(T=1) = +0.22 \pm 0.45 \pm 0.39
\end{equation}
in contrast with the calculated value \cite{mus00} $G_A^e (T=1) = -0.83 \pm
0.26$. This may be an indication that the anapole and other radiative correction effects in the nucleon
are somewhat larger (by a factor of 2-3) 
than expected based on these calculations. 

One should note that the calculation of $G_A^e (T=1)$
combined with the typical theoretical prediction $G_M^s = -0.3$ is substantially at variance with
the experimental result. Thus the SAMPLE experiment provides important new information
on the electroweak and flavor structure of the nucleon.

The second experiment \cite{HAPPEX2} is the HAPPEX experiment at the Jefferson Laboratory, which measures
the forward angle asymmetry from the proton at $Q^2 = 0.477$ (GeV/$c)^2$. Their quoted result
for the combination of form factors contributing to their measurement is 
\begin{equation}
G_E^s + 0.392 G_M^s = 0.025 \pm 0.020 \pm 0.024 \> .
\end{equation}
This precise result also points to a possibly reduced role for strange quarks, is consistent with
a previous result from neutrino scattering \cite{garvey}, and also rules out several theoretical
model predictions.

Most recently, the A4 collaboration at the Mainz facility MAMI has completed a measurement
of the forward angle asymmetry at $Q^2 = 0.225$ (GeV/$c)^2$ \cite{pva4}, which is sensitive to
the linear combination
\begin{equation}
G_E^s + 0.21 G_M^s\ \ \ .
\end{equation}
One expects a value for this quantity to be forthcoming in the near future.


\section{Conclusion and Outlook}

Clearly the recent set of experimental results in parity-violating electron-nucleon scattering
have provided significant new constraints on the contributions of strange
quark-antiquark pairs to the electromagnetic structure of the nucleon. The interpretation
is theoretically clean, and therefore the results can be taken as quite definitive (in contrast
to other methods of studying $\bar s s$ contributions to the spin and mass
of the nucleon). This program has been extremely successful in 
this regard, and we can look forward to additional higher precision data from these experiments.

As can be seen from the existing forward angle measurements, it is essential to perform measurements
that enable separate determination of the electric and magnetic form factors as functions
of momentum transfer. In addition, as we have learned from SAMPLE,
determination of the axial form factor $G_A^e$ is also necessary throughout the range of $Q^2$.
The new ``$G0$'' experiment \cite{g0} to be performed at Jefferson Lab
will have the capability to perform a precise determination of all of these form factors
as functions of $Q^2$. So the search for
strangeness effects will be continued both with higher precision and at higher momentum
transfers.

The SAMPLE experiment has also focused attention on the interesting new topic of electroweak
corrections to the axial form factor, and the importance of anapole effects and higher order
terms in ``box diagrams''.  The significance of nucleon structure effects in these 
amplitudes presents a new and important challenge for theory, with relevant applications
to precision electroweak tests in beta decay and atomic parity violation.

\section{Acknowledgements}

 The authors would like to thank G.~T. Garvey for a careful reading of the manuscript and
useful suggestions. This work was supported by NSF grant PHY-0071856 and DOE contract no. DE-FG02-00ER41146.


\begin{thebibliography}{99}

\bibitem{bazarko} A. O. Bazarko, {\it et al.}, {\it Z. Phys}. C65, 189
(1995).
\bibitem{ji01} B.W. Filippone and X. Ji, Adv. in Nucl. Phys., {\bf 26}, 1 (2001) .
\bibitem{ellis} J. Ellis and R. Jaffe, Phys. Rev. {\bf D 9}, 1444 (1974); 
Phys. Rev. {\bf D 10}, 1669E (1974).

\bibitem{gasser}J.~Gasser, H.~Leutwyler,  and M.~E.~Sainio, 
{\it Phys. Lett}. B253 (1991) 163. 
\bibitem{kaplan}D. Kaplan and A. Manohar, {\it Nucl. Phys. B} 310, 527 (1988).

\bibitem{bmck89}R. D. McKeown, {\it Phys. Lett}. B219, 140 (1989).
\bibitem{beck89}D. H Beck, {\it Phys. Rev}. D39, 3248 (1989).


\bibitem{hasty}
R. Hasty, {\it et al.}, Science {\bf 290}, 2117 (2000).

\bibitem{spayde}D.~T.~Spayde {\it et al.}, Phys. Rev. Lett. {\bf 84},
1106 (2000).

\bibitem{HAPPEX2}
K. A. Aniol, {\it et al.}, Phys.Lett. B509 (2001) 211-216.

\bibitem{g0} Jefferson Lab experiment 00-006, D. Beck, spokesperson.

\bibitem{pva4} Mainz experiment PVA4, D. von Harrach, spokesperson;
F. Maas, contact person.

\bibitem{musolf94a}M. J. Musolf, {\it et al.,}, {\it Phys. Rep}.  239 1 (1994).

\bibitem{mus92a}M.J. Musolf and T. W. Donnelly, Nucl. Phys. {\bf A546}, 509 (1992); erratum {\it ibid},
{\bf A550}, 564 (E) (1992). 

\bibitem{musolf90}M. J. Musolf and B. R. Holstein, {\it Phys. Lett}. 
B242, 461 (1990).

\bibitem{comment91} E. J. Beise and R. D. McKeown, Comm. Nucl. Part. Phys. {\bf 20}, 105 (1991).

\bibitem{annrev01} D. H. Beck and R. D. McKeown, Ann. Rev. Nucl. Part. Phys. {\bf 51} (2001).

\bibitem{sirlin}W. J. Marciano and A. Sirlin, Phys. Reve. {\bf D22}, 2695 (1980); erratum
{\it ibid}, {\bf D31}, 213 (1985).

\bibitem{musolf91}M.J. Musolf and B.R. Holstein, {\it Phys. Rev.} D43, 2956 (1991).

\bibitem{zeldovich} I. Zel'dovich, JETP Lett. {\bf 33}, 1531 (1957).

\bibitem{musolfphd}M. J. Musolf, Ph.D. thesis, Princeton University, unpublished.

\bibitem{haxton}W. C. Haxton, E. M. Henley, and M. J. Musolf, Phys. Rev.
Lett. {\bf 63}, 949 (1989); W. Haxton, Science {\bf 275}, 1753 (1997);
W.C. Haxton, C.P. Liu, and M.J. Ramsey-Musolf, Phys. Rev. Lett. {\bf 86}, 5247 (2001); nucl-th/0109014.

\bibitem{weiman}C. S. Wood {\it et al.}, Science {\bf 275}, 1759 (1997).

\bibitem{marciano} W. J. Marciano and A. Sirlin, Phys. Rev. {\bf D27},
552 (1983); W. J. Marciano and A. Sirlin, Phys. Rev. {\bf D29},
75 (1984).

\bibitem{mus00} S.-L.~Zhu, {\it et al}.,  Phys. Rev. {\bf D62}, 033008 (2000).

\bibitem{bira1}C. M. Maekawa and U. van Kolck, Phys. Lett. {\bf B478}, 73 (2000);
C. M. Maekawa, J. S. Viega, and U. van Kolck, Phys. Lett. {\bf B488}, 167 (2000).

\bibitem{riska00} D.-O. Riska, Nucl. Phys. {\bf A678}, 79 (2000).

\bibitem{mus00b}M.J. Ramsey-Musolf, Phys. Rev. {\bf C60}, 015501 (1999). 

\bibitem{garvey92}G.~T.~Garvey, S.~Krewald, E.~Kolbe, and
K.~Langanke, 
{\it Phys. Lett}. B289,  249 (1992);
G.~T.~Garvey, E.~Kolbe, K.~Langanke, and S.~Krewald, 
{\it Phys. Rev}. C48, 1919-1925 (1993).

\bibitem{hemmert} T. R. Hemmert, U.-G. Meissner, and S. Steininger,
Phys. Lett. {\bf B437} 184 (1998).

\bibitem{hammer}H.-W. Hammer and M.J. Ramsey-Musolf, Phys. Rev. {\bf C60}, 045204 (1999);
S.J. Puglia, M.J. Ramsey-Musolf, and S.-L. Zhu, Phys. Rev. {\bf D63}, 034014 (2001).

\bibitem{garvey}G.~T.~Garvey, W.~C.~Louis, and D.~H.~White, 
{\it Phys. Rev}. C48, 761-765 (1993) .




\end{thebibliography}
\end{document}